\documentclass[prd,aps,showpacs,preprintnumbers,amsmath,amssymb]{revtex4}
\usepackage[dvips]{graphicx}
\usepackage{epsf}
\usepackage{amsmath}
\usepackage{amssymb}

\usepackage{graphicx}% Include figure files
\usepackage{dcolumn}% Align table columns on decimal point
\usepackage{bm}% bold math
\def\be{\begin{equation}}
\def\ee{\end{equation}}
\def\ba{\begin{eqnarray}}
\def\ea{\end{eqnarray}}

\def\ga{\mathrel{\raise.3ex\hbox{$>$\kern-.75em\lower1ex\hbox{$\sim$}}}}
\def\la{\mathrel{\raise.3ex\hbox{$<$\kern-.75em\lower1ex\hbox{$\sim$}}}}

\newcommand{\fr}[2]{\frac{#1}{#2}}

\newcommand{\ini}{{\rm ini}}
\newcommand{\BD}{{\rm BD}}
\newcommand{\m}{{\rm m}}
\newcommand{\T}{{\cal T}}

\begin{document}

\title{Anisotropy due to Brans-Dicke Theory}

\author{Seokcheon Lee$^{1}$}

%\smallskip
%\medskip

\affiliation{1) Institute of Physics, Academia Sinica, Taipei, Taiwan 11529, R.O.C.}

\begin{abstract}

It is well known that the viscous Bianchi type-I metric of the Kasner form is not able to describe an anisotropic universe, which satisfies the second law of thermodynamics and the
dominant energy condition in Einstein's theory of gravity. We
investigate this problem in Brans-Dicke theory of gravity using
the Bianchi type-I metric with the perfect fluid. We show
that it is possible to have the dominant energy condition and the
growth of entropy in this model. Also we apply this model to explain the anomaly concerning the low quadrupole amplitude of the angular power spectrum of the temperature anisotropy observed in WMAP data.

\end{abstract}
\maketitle

%%%%%%%%%%%%%%%%%%%%%%%%%%%%%%%%%%%%%%%%%%%%%%%%%%%%%%%%%%%%%%%%%%%%%%%%%
\section{Introduction}
\setcounter{equation}{0}
%%%%%%%%%%%%%%%%%%%%%%%%%%%%%%%%%%%%%%%%%%%%%%%%%%%%%%%%%%%%%%%%%%%%%%%%%

One of the most interesting and complete scalar tensor theory of
gravities is Brans-Dicke (BD) theory depending on the BD
parameter $\omega_{\rm{BD}}$ and the gravitational constant $G$ is
replaced with the reciprocal of a scalar field $\Phi$ \cite{BD}.

Based on the Einstein's gravity theory, it is known that a viscous Bianchi type-I metric of the Kasner form is not possible to describe an anisotropic universe model which satisfies both the dominant energy condition (DEC) and the second law of thermodynamics \cite{0004055,0011027}. This problem might be resolved in some specific scalar-tensor theories of gravity \cite{0011027}.

The one of possible applications of the anisotropic universe to the phenomena is the explanation of anomalous feature of low quadrupole moment of the cosmic microwave background (CMB) anisotropy. If the large scale spatial geometry of the universe has the plane symmetric with an eccentricity at the last scattering of order $10^{-2}$, then the quadrupole amplitude of the CMB can be reduced with respect to the value of the best fit $\Lambda$CDM model without affecting higher multipoles of the angular power spectrum of the temperature anisotropy \cite{Sanz,0606266,07081168}.

The Bianchi type-I solutions of BD theory including perfect fluid is also well studied \cite{Ruban,Johri}. In what follows, we analyze in detail the solutions of this model and investigate both the entropy growth and the DEC of it.

This paper is organized as follows. In the next section we investigate the solution of BD theory in a Bianchi type-I metric with perfect fluid. We analyze the various energy conditions and the growth of entropy of this model in section $3$. We also consider the phenomena and the possible application of this model to the low quadrupole problem of the CMB in section $4$. In section $5$ we reach our conclusions.

%%%%%%%%%%%%%%%%%%%%%%%%%%%%%%%%%%%%%%%%%%%%%%%%%%%%%%%%%%%%%%%%%%%%%%%%%
\section{Brans-Dicke Field Equations}
\setcounter{equation}{0}
%%%%%%%%%%%%%%%%%%%%%%%%%%%%%%%%%%%%%%%%%%%%%%%%%%%%%%%%%%%%%%%%%%%%%%%%%

We start with the Bianchi type-I metric where the line element is given by
\be ds^2 = - c^2 dt^2 + \sum_{i=1,2,3} a_{i}(t)^2 dx_{i}^2 \, ,
\label{ds} \ee where $a_{i}$ is the scale factor of each spatial direction. BD theory is described by the action
\be S = \int d^4 x \sqrt{-g} \Biggl[ \fr{1}{16 \pi} \Bigl( \Phi R
- \fr{\omega_{\rm{BD}}}{\Phi} \nabla_{\mu} \Phi \nabla^{\mu} \Phi
\Bigr) + {\cal L}_{\rm{fluid}} \Biggr] \, , \label{S} \ee where $\Phi$
is a scalar field, $\omega_{\rm{BD}}$ is the BD parameter, and
${\cal L}_{\rm{fluid}}$ is the Lagrangian of the ordinary matter
component. From the above action (\ref{S}), we can find the BD field equations
\ba R_{\mu\nu} - \fr{1}{2} g_{\mu\nu} R &=& \fr{8 \pi}{\Phi}
T_{\mu\nu}^{\rm{m}} + 8 \pi T_{\mu\nu}^{\rm{BD}} \, ,  \label{Gmunu} \\
\Box \Phi &=& g^{\mu\nu} \nabla_{\mu} \nabla_{\nu} \Phi = \fr{8
\pi}{2
\omega_{\rm{BD}} + 3} T^{\rm{m}} \, , \label{boxPhi} \\
T_{\mu\nu}^{\rm{BD}} &=& \fr{1}{8 \pi} \Biggl[
\fr{\omega_{\rm{BD}}}{\Phi^2} \Bigl(\nabla_{\mu} \Phi \nabla_{\nu}
\Phi - \fr{1}{2} g_{\mu\nu} \nabla_{\lambda} \Phi \nabla^{\lambda}
\Phi \Bigr) + \fr{1}{\Phi} \Bigl(\nabla_{\mu} \nabla_{\nu} \Phi -
g_{\mu\nu} \Box \Phi \Bigr) \Biggr] \, . \label{TBD} \ea The field
equations (\ref{Gmunu}) are explicitly given in terms of the metric (\ref{S}) \ba
\fr{\dot{a}_{1}}{a_1} \fr{\dot{a}_2}{a_2} + \fr{\dot{a}_2}{a_2}
\fr{\dot{a}_3}{a_3} + \fr{\dot{a}_3}{a_3} \fr{\dot{a}_1}{a_1} &=&
\fr{8 \pi \rho}{\Phi} + \fr{\omega_{\rm{BD}}}{2}
\fr{\dot{\Phi}^2}{\Phi^2} - \fr{\dot{V}}{V} \fr{\dot{\Phi}}{\Phi} \, ,
\label{G00} \\
\fr{\ddot{a}_{j}}{a_{j}} + \fr{\ddot{a}_{k}}{a_{k}} +
\fr{\dot{a}_{j}}{a_{j}} \fr{\dot{a}_{k}}{a_{k}} &=& - \fr{8 \pi
P}{\Phi} - \fr{\omega_{\rm{BD}}}{2} \fr{\dot{\Phi}^2}{\Phi^2} +
\fr{\dot{a}_{i}}{a_{i}}\fr{\dot{\Phi}}{\Phi} + \fr{\Box \Phi}{\Phi} \, , \label{Gii} \\
\fr{\ddot{\Phi}}{\Phi} + \fr{\dot{V}}{V} \fr{\dot{\Phi}}{\Phi} &=&
\fr{8 \pi (\rho - 3P)}{(2 \omega_{\rm{BD}} + 3) \Phi} \, ,
\label{Phieq} \ea where $V = a_{1}a_{2}a_3$. From the Bianchi
identity we get \be \rho^{\rm{m}} V = \rho_{\rm{ini}}^{\rm{m}} V_{\rm{ini}} \, ,
\label{rhoV} \ee where $\rho_{\rm{ini}}^{\rm{m}}$ and $V_{\rm{ini}}$ are the matter density
and the volume element at a given time $t_{\rm{ini}}$.

The shear tensor has the form \be \sigma_{\mu\nu} =
h^{\alpha}_{\mu} U_{(\alpha; \beta)} h^{\beta}_{\nu} - \fr{1}{3}
\theta h_{\mu\nu} \, , \label{sigmamunu} \ee where $U^{\mu}$ is the
four velocity, which satisfies $U^{\mu}U_{\mu} = -1$, $\theta =
U^{\mu}_{;\mu}$ is the scalar expansion where the semicolon means
the covariant derivative, and $h_{\mu\nu}$ is the projection
tensor defined as $h_{\mu\nu} = g_{\mu\nu} + U_{\mu} U_{\nu}$. We
can identify the scalar expansion $\theta$ as the mean Hubble
expansion rate $H$ by \be \theta = H = \fr{1}{3} \Biggl(
\fr{\dot{a}_1}{a_1} + \fr{\dot{a}_2}{a_2} + \fr{\dot{a}_3}{a_3}
\Biggr) = \fr{1}{3} \fr{\dot{V}}{V} \, . \label{H} \ee Now we can find
the shear tensors from the above definition (\ref{sigmamunu}) \be
\sigma_{ii} = \fr{1}{3} a_i^2 \Biggl( \fr{\dot{a}_j}{a_j} +
\fr{\dot{a}_k}{a_k} - 2 \fr{\dot{a}_i}{a_i} \Biggr) \, ,
\hspace{0.2in} \sigma_{00} = 0 \, . \label{sigmaii} \ee

Later we will investigate the effect of anisotropy expansion in
the CMB temperature anisotropy. Thus we are only interested in the matter dominated epoch
in this model. In this case, we can choose the initial time $t_{\rm{ini}}$
as $t_{\rm{eq}}$ when the matter and the radiation components have the equal
amount of the energy density. Thus, we can find the value
of $t_{\rm{ini}}$ from the current observation from the Wilkinson Microwave Anisotropy Probe (WMAP) \cite{WMAP5}

\be \fr{t_{\rm{ini}}}{t_0} \simeq \fr{a_{\rm{eq}}}{a_0} =
\fr{\Omega_{0}^{r}}{\Omega_{0}^{\rm{m}}} \simeq 1.9 \times 10^{-4} \, . \label{tini} \ee
It is well known that the above field equations
have analytic solution in this epoch \cite{Ruban,Johri}. Equations (\ref{Phieq}) and
(\ref{rhoV}) give the relations \ba \fr{V}{V_{\ini}} &=& \Biggl(
\fr{\Phi}{\Phi_{\ini}} \Biggr)^{3(\omega_{\rm{BD}} + 1)} \, , \label{V} \\
\fr{\rho^{\rm{m}}}{\rho_{\ini}^{\rm{m}}} &=& \Biggl( \fr{\Phi}{\Phi_{\ini}}
\Biggr)^{-3(\omega_{\rm{BD}} + 1)} \, . \label{rho} \ea By integrating
equation (\ref{Gii}) with taking into account the Eqs. (\ref{V})
and (\ref{rho}), we find \be \Phi = \Phi_{\ini} \Bigl[r(t^2 + 2 {\cal
T} t) \Bigr]^{1/(3 \omega_{\rm{BD}} + 4)} \, , \label{Phi} \ee where $r
= (4 \pi \rho_{\ini}^{\rm{m}} / \Phi_{\ini})[(3 \omega_{\rm{BD}} + 4)/(2
\omega_{\rm{BD}} + 3)]$ and ${\cal T}$ is an arbitrary constant with the dimension of time. We can find the limit on the value of ${\cal T}$ in section $4$ in order to be consistent with data.
The metric coefficients $a_i$ are evaluated from the equations
(\ref{Gii}), (\ref{V}), (\ref{rho}), and (\ref{Phi}) to give \be
a_{i}(t) = a_{i}(t_{\ini}) \Biggl[\fr{t}{t + 2 {\cal
T}}\Biggr]^{\xi_i}V^{1/3} \, , \label{ai} \ee where $a_i(t_{\ini})$ and
$\xi_i$ are arbitrary constants satisfying \be
a_1(t_{\ini})a_2(t_{\ini})a_3(t_{\ini}) = V_{\ini} = 1 \, , \hspace{0.2in} \xi_1 + \xi_2 +
\xi_3 = 0 \, .  \label{a0ixi} \ee  From the above equation
(\ref{ai}), we can understand the isotropy of the universe. When
the age of the universe is about the same order as ${\cal T}$, then we should have
the small $\xi$ in order not to have the huge anisotropy of the
expansion. However, as the universe gets older, this contribution
becomes smaller and we will see the almost isotropic universe independent of the value of $\xi$.

We can derive the energy density and the pressure of BD field from
the Eq. (\ref{TBD}) \ba \rho^{\rm{BD}} &=& \fr{1}{8 \pi} \Biggl[
\fr{\omega_{\rm{BD}}}{2} \fr{\dot{\Phi}^2}{\Phi^2} -
\fr{\dot{V}}{V} \fr{\dot{\Phi}}{\Phi} \Biggr] \, , \label{rhoBD} \\
P^{\rm{BD}}_{i} &=& \fr{1}{8 \pi} \Biggl[ \fr{\omega_{\rm{BD}}}{2}
\fr{\dot{\Phi}^2}{\Phi^2} + \fr{\ddot{\Phi}}{\Phi} + \fr{\dot{V}}{V}
\fr{\dot{\Phi}}{\Phi} - \fr{\dot{a}_i}{a_i} \fr{\dot{\Phi}}{\Phi}
\Biggr] \, . \label{PBDi} \ea %** {\bf One conceptional problem in the
%energy density of BD field is that $\rho^{\rm{BD}}$ is negative
%%\Biggl[ \fr{\omega_{\rm{BD}}}{2} - 3(\omega_{\rm{BD}} + 1) \Biggr]
%\fr{\dot{\Phi}^2}{\Phi^2} = \fr{-1}{16 \pi} (5 \omega_{\rm{BD}} +
%6) \fr{\dot{\Phi}^2}{\Phi^2} \ee }**
From these equations (\ref{rhoBD}) and (\ref{PBDi}), we can define the directional components of the equation of state (eos) of BD field \be P_{i}^{\rm{BD}} \equiv \lambda_{i} \rho^{\rm{BD}} \, ,
\label{PBDi2} \ee where $\lambda_i$ is given by  \be \lambda_i = \fr{1}{5
\omega_{\rm{BD}} + 6} \Biggl[ \omega_{\rm{BD}} + 2 - (3
\omega_{\rm{BD}} + 4) \fr{t^2 + 2 \T t}{(t + \T)^2} + 2 \xi_i
\T \fr{3 \omega_{\rm{BD}} + 4}{t + \T} \Biggr] \, . \label{lambdai}
\ee It will be useful to define the sum of the pressure parameters
by \be \Delta \equiv \sum_{i} \lambda_{i} = \fr{3}{5
\omega_{\rm{BD}} + 6} \Biggl[ \omega_{\rm{BD}} + 2 - (3
\omega_{\rm{BD}} + 4) \fr{t^2 + 2 \T t}{(t + \T)^2} \Biggr] \, .
\label{Delta} \ee

%%%%%%%%%%%%%%%%%%%%%%%%%%%%%%%%%%%%%%%%%%%%%%%%%%%%%%%%%%%%%%%%%%%
\vspace{1in}
\begin{widetext}
\begin{figure}[htp]
\centering
\includegraphics[scale=0.3]{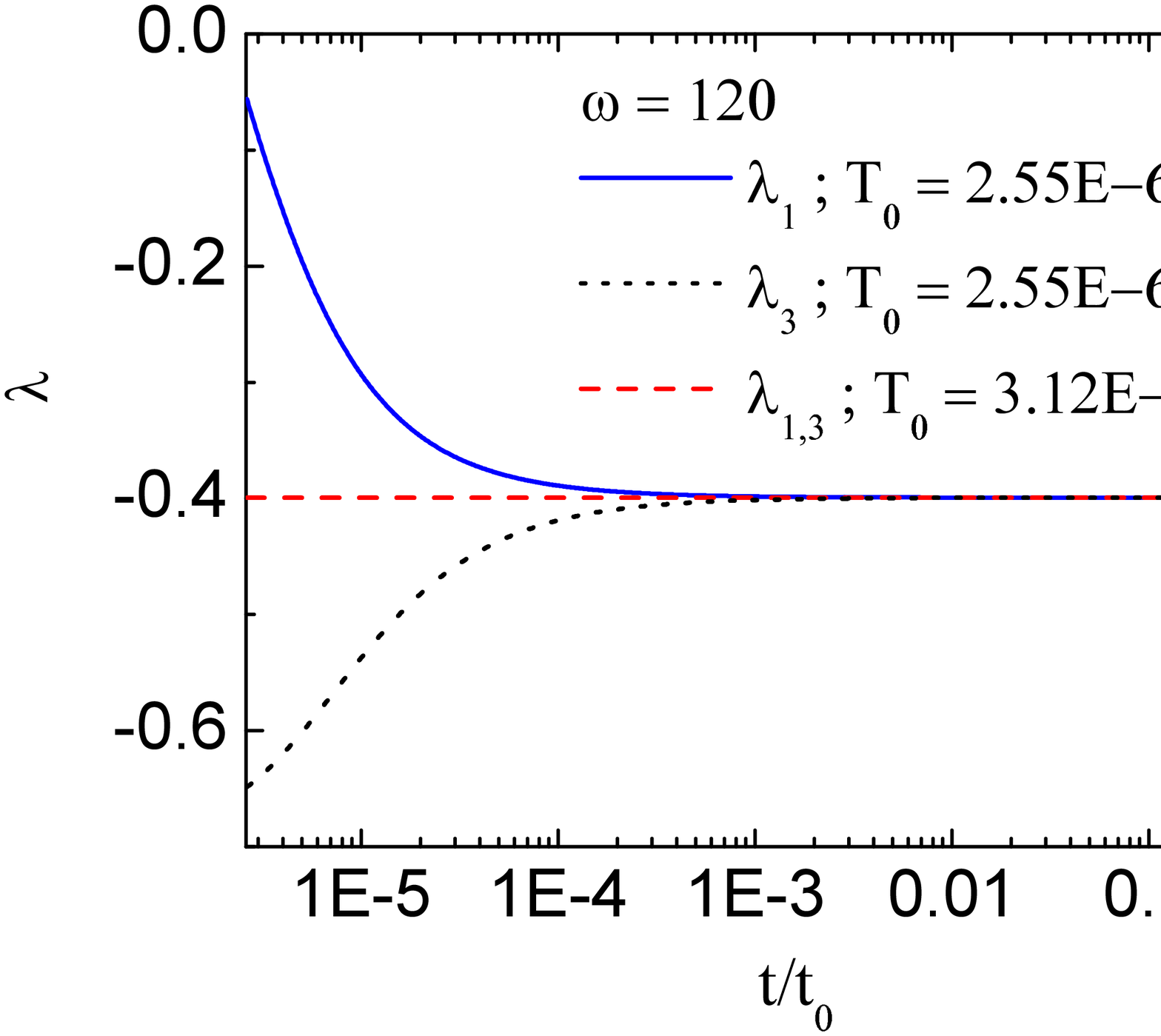}\hfill\includegraphics[scale=0.3]{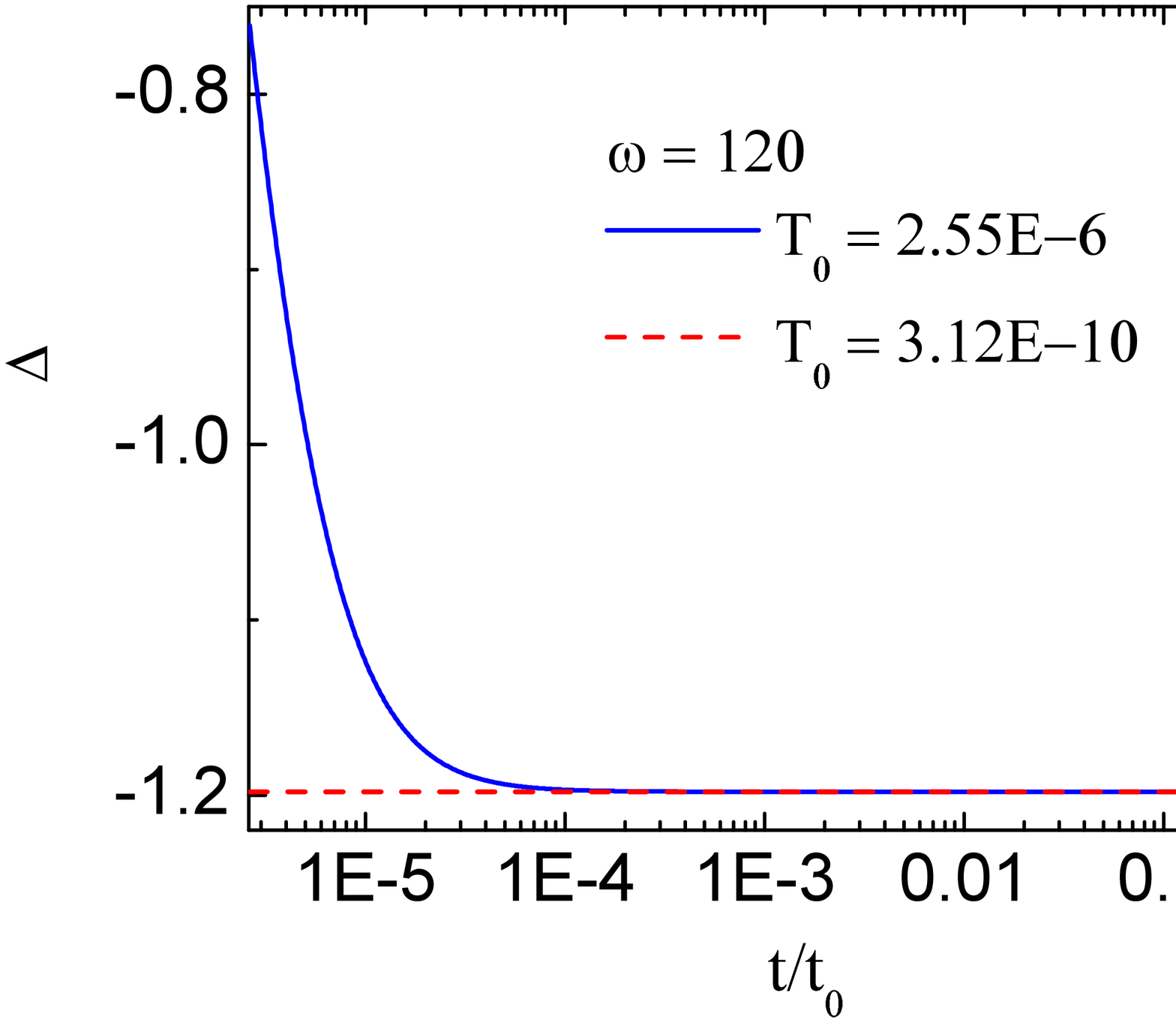}
%\includegraphics[scale=0.4]{D.eps}\\
%\includegraphics[width=0.4\textwidth,viewport=50 260 400
%1000,clip]{l.eps}
%\hfill
%\includegraphics[width=0.4\textwidth,viewport=50 260 400 1000,clip]{D.eps}\\
\vspace{-1.5cm}
\caption{The directional components ($\lambda_{i}$s) and sum of them ($\Delta$) of the eos of BD for the different values of ${\cal T}$ when we use choose the value of $\omega_{\BD}$ as $120$. a) The directional components of the eos of BD field ($\lambda_{i}$) for the two specific values of ${\cal T}$. When ${\cal T} = 3.12 \times 10^{-10} t_{0}$, all of $\lambda_{i}$s converge to the one value $-0.4$ during the interested epoch. b) The sum of $\lambda_{i}$ indeed converge to $-1.2$ when ${\cal T} = 3.12 \times 10^{-10} t_{0}$.}\label{fig1}
\end{figure}
\end{widetext}
%%%%%%%%%%%%%%%%%%%%%%%%%%%%%%%%%%%%%%%%%%%%%%%%%%%%%%%%%%%%%%%%%%%%%%%%%%%

In Fig \ref{fig1}, we show the evolutions of $\lambda_{i}$s and $\Delta$ for $\omega_{\BD} = 120$ which is chosen by the CMB power spectrum constrain for $\omega_{\rm{BD}}$ ($> 120$) in the matter dominated epoch \cite{Acquaviva}. We choose
$t_0$ as the present age of the universe and $\T$ as $3.12 \time2 10^{-10} t_0$ or $2.5 \times 10^{-6} t_0$ in this figure. The first choice of $\T$ value is given to be consistent with data and the second one is for the demonstration for $\T$ with the similar magnitude as $t_{\rm{ini}}$. We also specify the planar symmetry universe ({\it i.e.} $a_{1} = a_{2} = a$ and $a_{3} = b$) in this figure, which constrains the value of $\xi_{i}$s as shown in the next section. The directional components of the eos of BD field ($\lambda_{i}$s) converge to one finite value ($-0.4$) independent with the value of $\omega_{\BD}$ when the time increases. Thus, the sum of each component of eos which is identical to the eos of BD field ($\Delta$) also converges to the finite value ($-1.2$) near the present. These are shown in the above figure. In the left panel of the figure \ref{fig1}, the solid and dotted lines show the evolutions of $\lambda_{1}$ (same as $\lambda_{2}$) and $\lambda_{3}$ when $\T = 2.55 \times 10^{-6} t_0$. In this case we can see the anisotropic pressure contributions on each direction. However, dashed line indicates the both $\lambda_{1}$ and $\lambda_{3}$ when $\T = 3.12 \times 10^{-10} t_0$. They show the identical evolution in this case and the anisotropy of the universe is very small. The right panel of the above figure shows the evolution of $\Delta$ for the different values of $\T$. We can find two interesting features from this figure. The first, in the matter dominated epoch all of the pressure contribution comes only from the BD field because the pressure of matter is zero. So anisotropy will be decreased because each directional dependent pressure components of BD field converge to the same value. The other is that the energy density of BD field is increased as the universe expanded ($\rho^{BD} \propto a^{0.6}$) instead of decreasing because the effective eos of BD field is smaller than $-1$.

%%%%%%%%%%%%%%%%%%%%%%%%%%%%%%%%%%%%%%%%%%%%%%%%%%%%%%%%%%%%%%%%%%%%%%%%%
\section{Energy Conditions and Thermodynamics}
\setcounter{equation}{0}
%%%%%%%%%%%%%%%%%%%%%%%%%%%%%%%%%%%%%%%%%%%%%%%%%%%%%%%%%%%%%%%%%%%%%%%%%

The energy-momentum conservation of total energy ($\nabla_{\mu} \Bigl[ T^{\mu
\rm{m}}_{\nu} / \Phi + T^{\mu \rm{BD}}_{\nu} \Bigr] = 0$) is a consistency condition originating from
the geometric Bianchi identity ($\nabla_{\mu} G^{\mu}_{\nu} = 0$) with
$G^{\mu}_{\nu}$ being the Einstein tensor given in Eq.
(\ref{Gmunu}). If we use equations (\ref{TBD}), (\ref{rhoBD}), and
(\ref{PBDi}), then we have \be \nabla_{\mu} \Biggl[ \fr{T^{\mu \rm{m}}_{\nu}}{\Phi}
+ T^{\mu \rm{BD}}_{\nu} \Biggr]
= \fr{\dot{\rho}^{\rm{m}}}{\Phi} - \fr{\rho^{\rm{m}}}{\Phi} \fr{\dot{\Phi}}{\Phi} + \fr{\rho^{\rm{m}}}{\Phi} \fr{\dot{V}}{V} + \dot{\rho}^{\rm{BD}} + \fr{\dot{V}}{V} \rho^{\rm{BD}} +
\sum_{i=1,2,3} \fr{\dot{a}_i}{a_i} P^{\rm{BD}}_{i} = 0 \, .
\label{nablaTBD} \ee Now we can express several useful quantities
explicitly by using the equations from (\ref{V}) to (\ref{ai}) \ba
\fr{\dot{\Phi}}{\Phi} &=& \fr{2}{(3 \omega_{\rm{BD}} + 4)}
\fr{t + {\cal T}}{t^2 + 2 {\cal T} t} \, , \label{dotPhi} \\
\fr{\dot{V}}{V} &=& 3 (\omega_{\rm{BD}} + 1) \fr{\dot{\Phi}}{\Phi} \, ,
\label{dotV} \\ \fr{\ddot{\Phi}}{\Phi} &=& - \fr{\dot{V}}{V}
\fr{\dot{\Phi}}{\Phi} + \fr{1}{t + {\cal T}} \fr{\dot{\Phi}}{\Phi} \, ,
\label{ddotPhi} \\ \fr{\dot{a}_i}{a_i} &=& \Biggl[ \xi_i {\cal T}
\fr{(3 \omega_{\rm{BD}} + 4)}{t + {\cal T}} + (\omega_{\rm{BD}} +
1) \Biggr] \fr{\dot{\Phi}}{\Phi} \, . \label{dotai} \ea After we use
the above equations (\ref{dotPhi}) - (\ref{dotai}) into the
Bianchi identity (\ref{nablaTBD}), we have \be 8 \pi \nabla_{\mu} \Biggl[ \fr{T^{\mu \rm{m}}_{\nu}}{\Phi}
+ T^{\mu \rm{BD}}_{\nu} \Biggr] = \fr{(2 \omega_{\rm{BD}} + 3)}{(t + {\cal T}) (t^2
+ 2 {\cal T} t)} \fr{\dot{\Phi}^2}{\Phi^2} \Biggl[ 2 {\cal T}^2  \Bigr( 1 - \sum_{i=1,2,3} \xi_{i}^2 \fr{( 3
\omega_{\rm{BD}} + 4)}{(2 \omega_{\rm{BD}} + 3)} \Bigr) \Biggr] =
0 \, . \label{nablaTBD2} \ee To satisfy the above equation, we have \ba
\omega_{\rm{BD}} &=& -\fr{3}{2}  \hspace{0.2in} {\rm{or}} , \label{omegaBD} \\ \sum_{i =
1,2,3} \xi_{i}^2 &=& \xi_1^2 + \xi_2^2 + \xi_3^2 = \fr{2 \omega_{\rm{BD}} + 3}{3\omega_{\rm{BD}} + 4} \, . \label{xi2} \ea Again when we adopt the CMB power spectrum constrain for $\omega_{\rm{BD}}$ ($> 120$) in the matter dominated epoch, then we can abort the first condition in the above equation (\ref{omegaBD}). If we consider the ellipsoidal universe ({\it i.e.} $a_{1} = a_{2} = a$ and $a_{3} = b$), then we can find the values of $\xi_{i}$ from the equations (\ref{a0ixi}) and (\ref{xi2})  \be \xi_1 = \xi_2 = \pm \sqrt{\fr{2 \omega_{\BD} + 3}{6 (3 \omega_{\BD} + 4)}} \, , \hspace{0.2in} \xi_3 = - 2 \xi_{1} = \mp 2 \sqrt{\fr{2 \omega_{\BD} + 3}{6 (3 \omega_{\BD} + 4)}} \, . \label{xi123} \ee

\vspace{2cm}

\begin{widetext}
%%%%%%%%%%%
\begin{figure}[htp]
\centering
\includegraphics[scale=0.3]{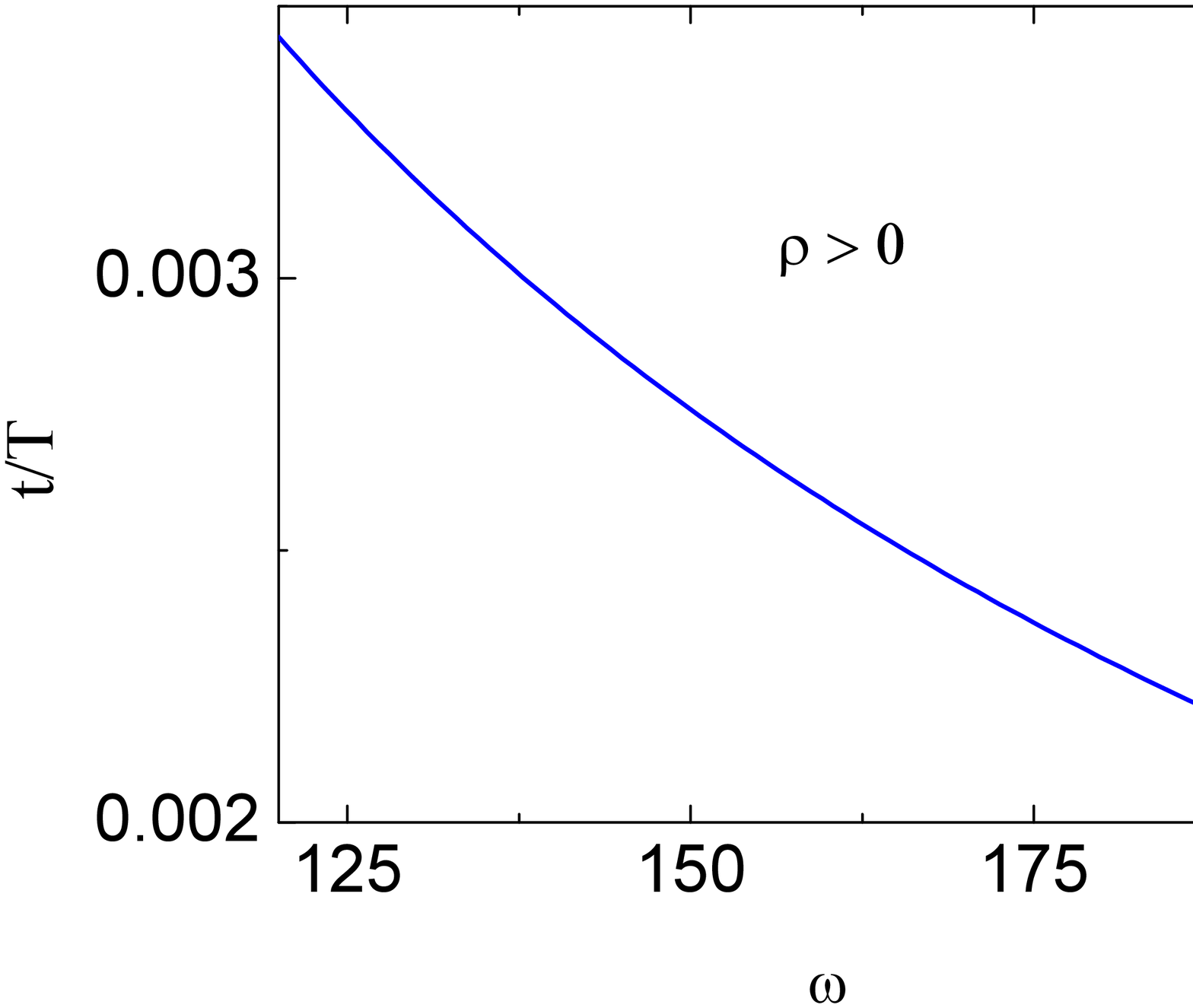}
\hfill
\includegraphics[scale=0.3]{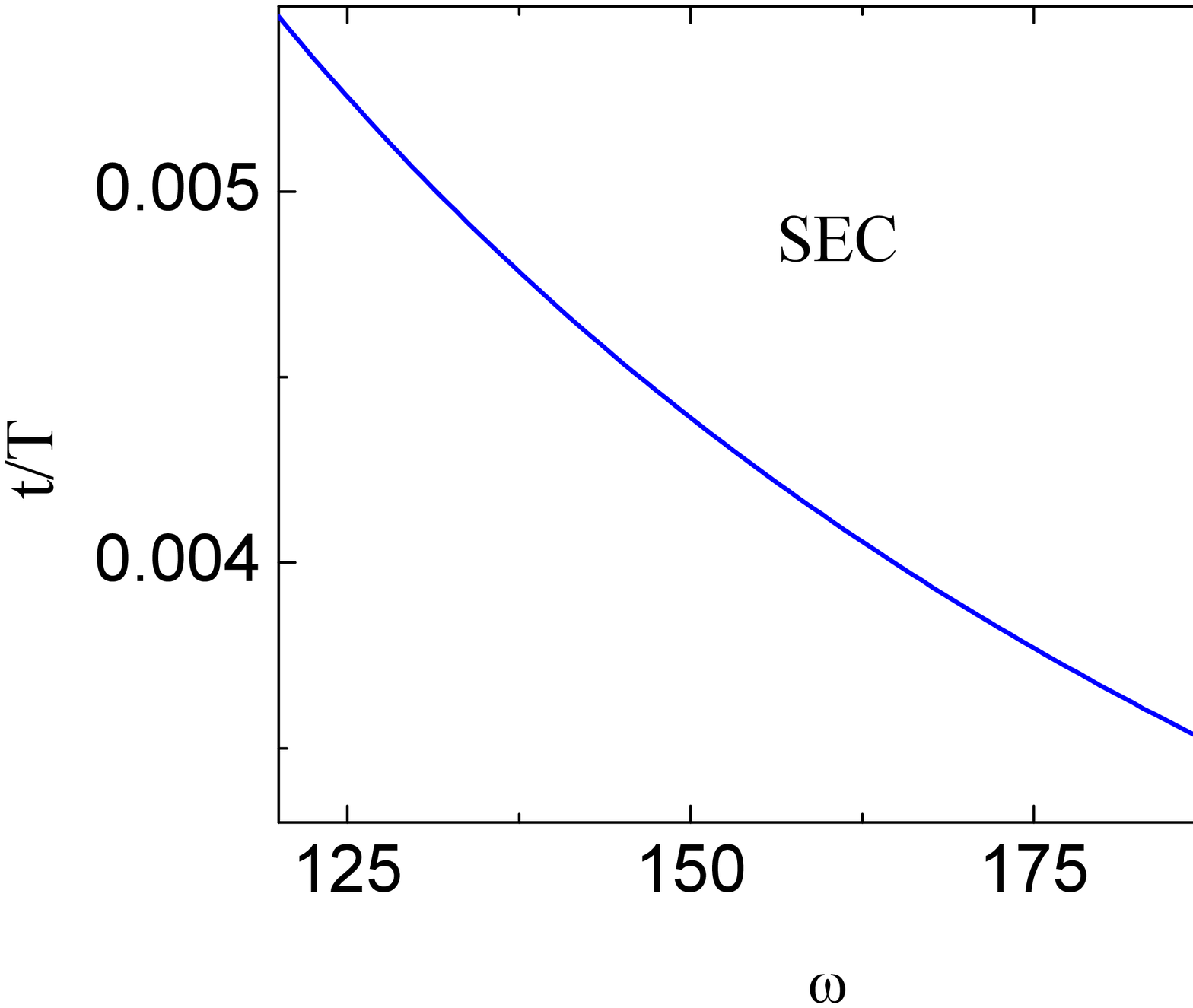}
\vspace{-2cm}
\caption{The values of coefficients of equations (\ref{trhopos}) and (\ref{SEC}) when we use the constraint $\omega_{\BD} > 120$. a) To get the positive energy density, the investigated time should be bigger than $0.0034 {\cal T}$.  b) The SEC can be satisfied when the observing time is bigger than $0.0055 {\cal T}$.} \label{fig2}
\end{figure}
%%%%%%%%%%%%%%
\end{widetext}

The total energy density, the pressures, and the sum of pressures can be explicitly represented by \ba
\rho_{\rm{total}} &=& \rho^{\rm{BD}} + \fr{\rho^{\m}}{\Phi} = \fr{1}{8 \pi}
\fr{\dot{\Phi}^2}{\Phi^2} \Biggl[3(\omega_{\BD} + 1)^2 - \fr{(6 \omega_{\BD}^2 + 17 \omega_{\BD} +12) \T^2}{2 (t + {\cal T})^2} \Biggr] \, , \label{rhoBD2}  \\
P^{\rm{BD}}_i &=& \fr{1}{8 \pi} \fr{\dot{\Phi}^2}{\Phi^2} \Biggl[(\omega_{\BD} + 1) - \fr{3 \omega_{\BD} + 4}{2} \fr{\Bigl(2 \xi_{i} \T t + (2 \xi_{i} + 1) \T^2 \Bigr)}{(t + \T)^2} \Biggr] \, ,  \label{PBD2} \\
\rho^{\rm{BD}} &=& \sum_{i} P^{\rm{BD}}_i =  \fr{1}{8 \pi} \fr{\dot{\Phi}^2}{\Phi^2} \Biggl[ 3 (\omega_{\BD} + 1) - \fr{( 9 \omega_{\BD} + 12)}{2} \fr{\T^2}{(t + \T)^2} \Biggr] \, . \label{PBDsum} \ea From the above equations (\ref{rhoBD2}) - (\ref{PBDsum}) we can find the constraints of $t$ to get the positive definite of energy density, dominant energy condition (DEC), and strong energy condition (SEC)
\ba t_{\rho \geq 0} &\geq& \Biggl[-1 + \sqrt{\fr{ (2 \omega_{\BD} + 3)(3 \omega_{\BD} + 4)}{6 (1 + \omega_{\BD})^2}} \, \Biggr] \T \, , \label{trhopos} \\ t_{\rm{DEC}} &\geq& \Biggl[ \fr{(-2 -2 \omega_{\BD} + \xi_{i}) + \sqrt{2 + 2 \omega_{\BD} + \xi_{i}^2}}{2 (1 + \omega_{\BD})} \, \Biggr] \T \, , \label{DEC} \\ t_{\rm{SEC}} &\geq& \Biggl[ - 1 + \sqrt{\fr{(3 \omega_{\BD} + 4)(\omega_{\BD} + 3)}{3 (\omega_{\BD} + 1) (\omega_{\BD} + 2)}} \, \Biggr] \T \, . \label{SEC} \ea

As we can see in the left panel of the figure \ref{fig2}, if we restrict $\omega_{\BD} > 120$, then we can satisfy the positive energy density constraint when the observing time is greater than $0.0034 {\cal T}$. The DEC is always satisfied for any time interval. The SEC is contented with $t > 0.0055 {\cal T}$.

In general the energy momentum tensor for a viscous fluid is given by \be T_{\mu\nu} = \Bigl[\rho + (p - \epsilon \theta) \Bigr] U_{\mu} U_{\nu} + (p - \epsilon \theta) g_{\mu\nu} - 2 \eta \sigma_{\mu\nu} \label{Tvis} \ee where $\epsilon$ and $\eta$ are the bulk and shear viscosities, respectively. If we use the equation (\ref{PBDi}) and define $P^{\rm{BD}}_{i} = p_{i} - 2 \eta \sigma^{i}_{i}$ (no sum over $i$) then we can find the shear viscosity as \be \eta = \fr{3}{8 \pi} \Biggl[ \fr{\xi_{i}}{\xi_{j} + \xi_{k} - 2 \xi_{i}} \fr{3 \omega_{\BD} + 4}{t + {\cal T}} + \fr{\omega_{\BD} + 1}{(\xi_{j} + \xi_{k} - 2 \xi_{i}) {\cal T}} \Biggr] \fr{1}{t^2 + 2 {\cal T} t} \label{eta} \ee The second law of thermodynamics requires $\eta > 0$ and this condition is satisfied by given $\xi$s (\ref{xi123}) and ${\cal T} = 3.12 \times 10^{-10} t_0$. Thus this model satisfies both various energy conditions and the second law of thermodynamics.

%%%%%%%%%%%%%%%%%%%%%%%%%%%%%%%%%%%%%%%%%%%%%%%%%%%%%%%%%%%%%%%%%%%%%%%%%
\section{Phenomena}
\setcounter{equation}{0}
%%%%%%%%%%%%%%%%%%%%%%%%%%%%%%%%%%%%%%%%%%%%%%%%%%%%%%%%%%%%%%%%%%%%%%%%%

\vspace{2.0cm}
\begin{widetext}
%%%%%%%%%%%
\begin{figure}[htp]
\centering
\includegraphics[scale=0.3]{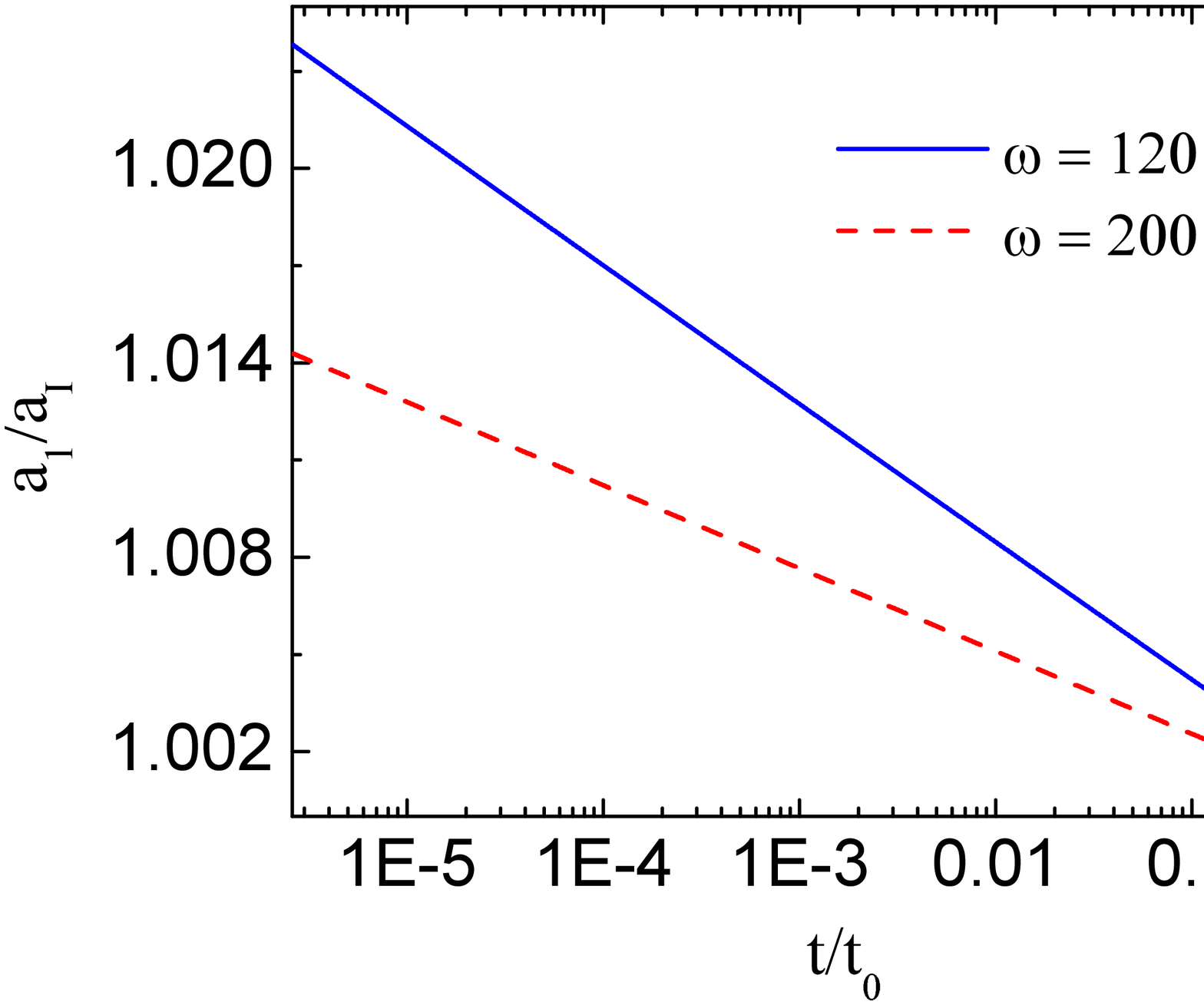}
\hfill
\includegraphics[scale=0.3]{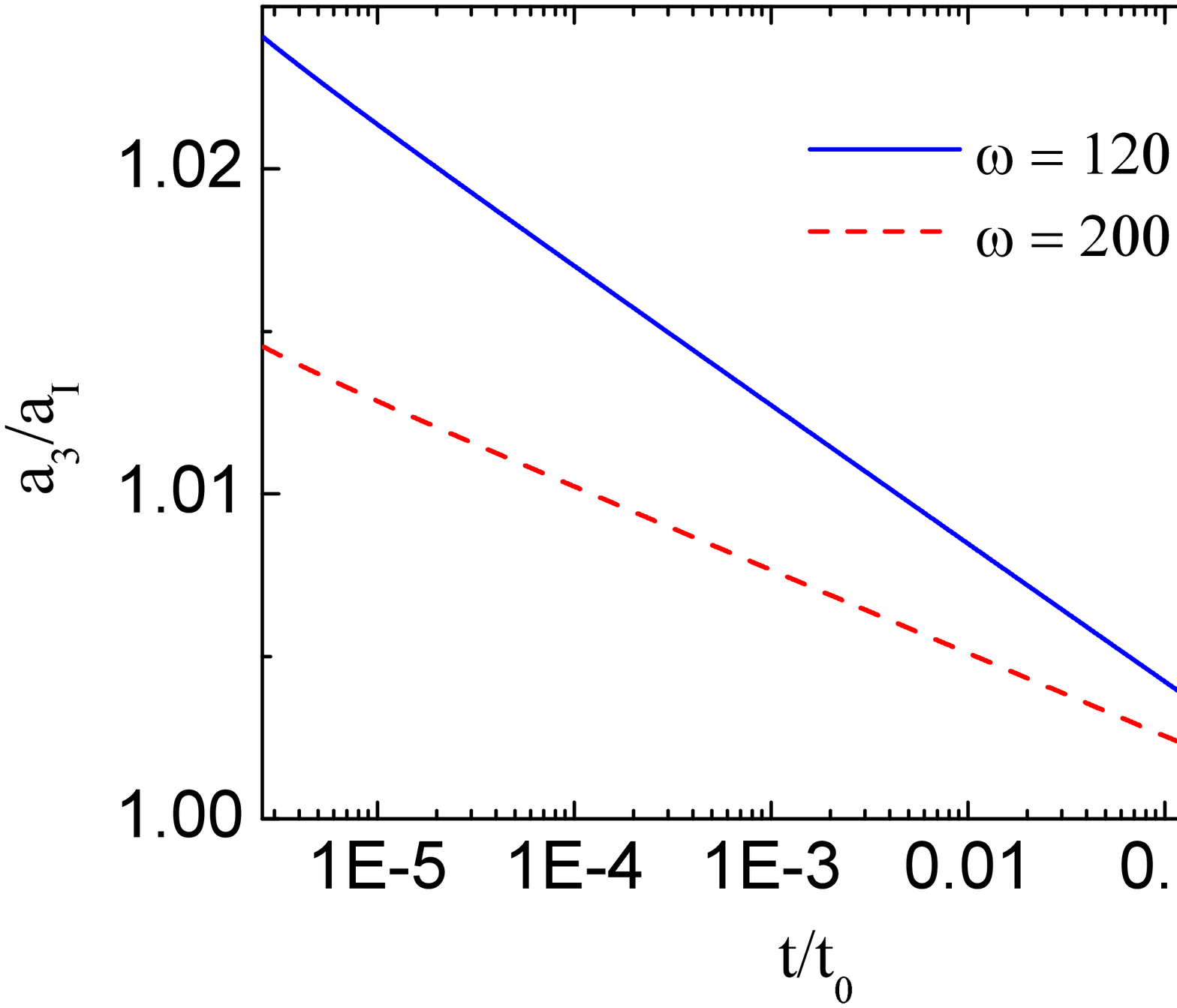}
\vspace{-2cm}
\caption{The time evolution of the ratio $a_{i}(t)$ to $a_{\rm{I}}(t)$ for the different values of $\omega_{\rm{BD}}$ with ${\cal T} = 3.12 \times 10^{-10} t_0$. The anisotropies on each direction get smaller as the universe evolves.  a) The evolution of $a_{1}(t)/ a_{\rm{I}}(t)$ for $\omega_{\BD} = 120$ (solid line) and $200$ (dashed line). b) $a_{3}(t)/ a_{\rm{I}}(t)$ for the same values of $\omega_{\BD}$ with same notation as in the left panel.} \label{fig3}
\end{figure}
%%%%%%%%%%%%%%
\end{widetext}

In this section we investigate the cosmological models with planar symmetry. In this case, we have the relation between $\xi_{i}$s as given by the equation (\ref{xi123}). If we check the scale factor in this epoch, then from the equation
(\ref{ai}) we have \be \fr{a_i(t)}{a_i(t_0)} = \Biggl[ \fr{\tau + 2 {\cal T}_0 \tau}{\tau +
2 {\cal T}_0} \Biggr]^{\xi_i} \Biggl [\fr{\tau^2 + 2 {\cal T}_0 \tau}{1 + 2 {\cal T}_0} \Biggr]^{(\omega_{\BD} + 1)/(3 \omega_{\BD} + 4)} \, , \label{ai2} \ee where $\tau = t/t_{0}$, ${\cal T}_{0} = {\cal T} / t_{0}$, and $t_0$ is again the present age of the universe. If the universe is isotropic and matter dominated, then the scale factor evolves as $a_{I}(t) = a_{I}(t_0) (t/t_{0})^{3/2}$. If we set $a_{I}(t_0) = a_{i}(t_0) = 1$, then we can show the ratio between $a_{i}(t)$ and $a_{I}(t)$ as in the figure \ref{fig3}. We show the ratio of $a_{1}(t)/a_{I}(t)$ in the left panel of the figure \ref{fig3}. We choose ${\cal T}_0 = 3.12 \times 10^{-10}$ in this figure same as in the previous section. As we expected, the anisotropy of the universe is getting smaller as time evolves and it will become isotropic at near present. For the bigger values of $\omega_{\BD}$, we have smaller anisotropies. In the right panel of the figure \ref{fig3}, we also show the ratio of $a_{3}(t) / a_{I}(t)$.

\vspace{2cm}
\begin{widetext}
%%%%%%%%%%%
\begin{figure}[htp]
\centering
\includegraphics[scale=0.3]{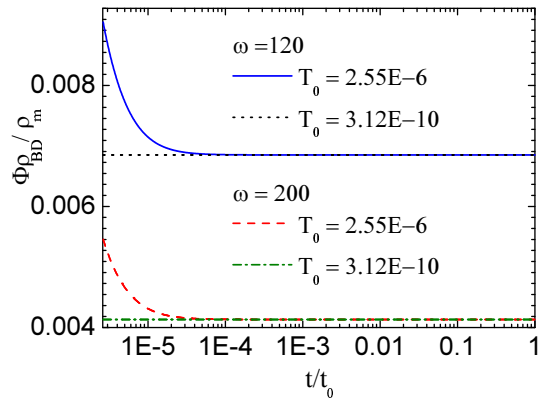}
\vspace{-2cm}
\caption{The cosmological evolution of ratio $\rho^{\BD}$ to $\rho^{\rm{m}} / \Phi$ for the different values of $\omega_{\BD}$s with ${\cal T} = 2.55 \times 10^{-6} t_{0}$ and $3.12 \times 10^{-10} t_{0}$. When ${\cal T} = 3.12 \times  10^{-10} t_0$, the value of $\rho^{\BD} \Phi / \rho^{\rm{m}}$ converges to $5 / (6 \omega_{\BD})$ as we can find in the equation (\ref{rho2}).} \label{fig4}
\end{figure}
%%%%%%%%%%%%%%
\end{widetext}

If we check the evolution of the energy density of matter and BD field,
then from the equations (\ref{rho}) and (\ref{rhoBD}) we have the absolute values of the ratio of them \be
\Biggl| \fr{ \Phi \rho^{\BD}}{\rho^{m}} \Biggr| = \fr{5 \omega_{\BD} + 6}{(2 \omega_{\BD})(3 \omega_{\BD} + 4)} \fr{(\tau + {\cal T}_0)^2}{\tau^2 + 2 {\cal T}_0 \tau} \, . \label{rho2} \ee As we can see in the figure \ref{fig4}, the absolute value of the energy density of BD field is decreased as time goes compared to that of matter. From the equation (\ref{ai}), we can see that
as time goes the universe will be more isotropic and we will
hardly see anisotropy. If the matter field keeps the dominant component and the universe
becomes isotropic, then we have the matter dominated epoch with
scaling $a \sim t^{2/3}$. The Brans-Dicke field
dominated epoch in the isotropic universe is
investigated intensively in the literature \cite{HKim}. In this
case, the universe undergoes zero acceleration epoch $a \sim t$.

If we assume that the universe is plane symmetric with an eccentricity at the last scattering surface ($e_{\rm{dec}}$) of order $10^{-2}$, then the quadrupole amplitude of the CMB temperature fluctuation can be reduced with respect to the observed  value of the WMAP \cite{08030732} without affecting the amplitudes of higher multipoles \cite{0606266}. The temperature anisotropy measured in a given direction of the sky can be expanded in spherical harmonics as \be \fr{\Delta T (\hat{n})}{T_{0}} = \sum_{l,m} a_{lm} Y_{lm}(\hat{n}) \, , \label{DeltaT} \ee where $\hat{n}$ is the direction of the photon momentum, $T_0 \simeq 2.73 K$ is the present average temperature of CMB radiation, and $a_{lm}$ are the multipoles. From this, the power spectrum is given by \be \fr{\Delta T_{l}}{T_0} = \sqrt{\fr{1}{2\pi} \fr{l (l+1)}{2l +1} \sum_{m} |a_{lm}|^2} \, . \label{Tl} \ee In the plane symmetric universe with a small eccentricity, the temperature anisotropy is a linear superposition of the isotropic temperature fluctuation $\Delta T_{I}$ and fluctuations due to the anisotropic background $\Delta T_{A}$ \cite{9605123} \be \Delta T = \Delta T_{I} + \Delta T_{A} \, . \label {DeltaT2} \ee We may write the amplitude \be a_{lm} = a_{lm}^{I} + a_{lm}^{A} \, . \label{alm} \ee From the null geodesic equation, a photon emitted at the last scattering surface with temperature $T_{\rm{dec}}$ reaches to the observer with the temperature \be T_{0}(\hat{n}) = T_{0} (1 - \fr{1}{2} e_{\rm{dec}}^2 n_{3}^2) \, , \label{T0n} \ee where $T_{0} = T_{\rm{dec}} a_{\rm{dec}}$ and $e_{\rm{dec}} = \sqrt{ \Bigl( \fr{a_{\rm{dec}}}{b_{\rm{dec}}} \Bigr)^2 -1}$. Therefore, the temperature anisotropy due to the anisotropic background is given by \be \fr{\Delta T_{A}}{T_0} = \fr{1}{2} e_{\rm{dec}}^2 n_{3}^2 \, . \label{DeltaTA} \ee We need to get $e_{\rm{dec}} \simeq 0.67 \times 10^{-2}$ in order to get the proper mean value of observed quadrupole anisotropy $(\Delta T_{A})^2|_{\rm{mean}} \simeq 427.3 \mu K^2$.

\vspace{2.0cm}
\begin{widetext}
%%%%%%%%%%%
\begin{figure}[htp]
\centering
\includegraphics[scale=0.3]{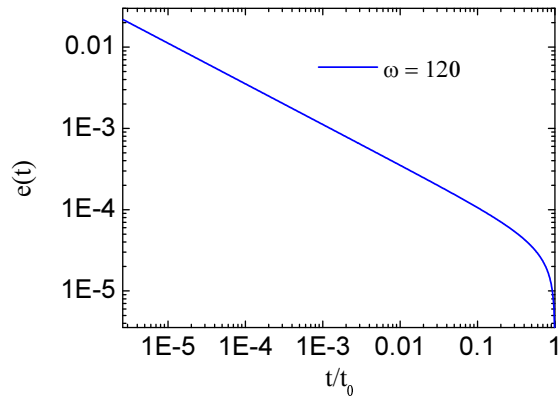}
\vspace{-2cm}
\caption{The cosmological evolution of $e(t)$ for ${\cal T} = 3.12 \times 10^{-10} t_0$ with $\omega_{\BD} = 120$.} \label{fig5}
\end{figure}
%%%%%%%%%%%%%%
\end{widetext}

In the figure \ref{fig5} we show the cosmological evolution of the eccentricity of the model (e) for the $\omega_{\BD} =120$. As we show in the figure when $\T = 3.12 \times 10^{-10} t_0$ we can obtain the expected value of $e_{\rm{dec}} \simeq 0.67 \times 10^{-2}$.

%%%%%%%%%%%%%%%%%%%%%%%%%%%%%%%%%%%%%%%%%%%%%%%%%%%%%%%%%%%%%%%%%%%%%%%%%
\section{Conclusion}
\setcounter{equation}{0}
%%%%%%%%%%%%%%%%%%%%%%%%%%%%%%%%%%%%%%%%%%%%%%%%%%%%%%%%%%%%%%%%%%%%%%%%%

We show that the Bianchi type-I metric in Brans-Dicke theory of gravity with matter fluid can satisfy the second law of thermodynamics and the various energy conditions. The cosmological evolution of the Brans-Dicke field reduces the possible primordial anisotropy of the universe.

We have considered the alternative explanation for the anomaly in the quadrupole amplitude of CMB temperature spectrum. A small eccentricity due to the Brans-Dicke field may solve this conundrum.

%\begin{acknowledgments}

%\end{acknowledgments}

%%%%%%%%%%%%%%%%%%%%%%%%%%%%%%%%%%%%%%%%%%%%%%%%%%%%%%%%%%%%%%%%%%%%%%%

\end{document}